\documentclass[aps,nofootinbib,reprint,nobibnotes,showpacs,superscriptaddress]{revtex4-1}

\usepackage[english]{babel}
	\addto\captionsenglish{%
	}

\usepackage{graphicx}

	\graphicspath{{graphics/}}

\usepackage{amsmath}
\usepackage{amssymb}
\usepackage[version=3]{mhchem}	% chemical formulas -> e.g. \ce{^{130}Te} instead of $^{130}$Te
\usepackage{bm}	% bold
\usepackage{nicefrac}	% fractions: a/b
\usepackage{mathrsfs}

\usepackage[hidelinks,linktocpage]{hyperref}
\hypersetup{
	colorlinks,
	linkcolor={red!70!black},
	citecolor={green!70!black},
	urlcolor={blue!70!black}
}
\usepackage[usenames,dvipsnames]{xcolor}	%colored text
\usepackage{comment}	% commented (non considered) environment \begin{comment} ... \end{comment}
\usepackage{upgreek}	% non italic μ symbol

\newcommand{\bb}{$0\nu \beta \beta$} % neutrinoless double beta decay
\newcommand{\mbb}{m_{\beta \beta}} % m_ee
\newcommand{\mbbM}{\mbb^{\max}} % m_ee max
\newcommand{\mbbm}{\mbb^{\min}} % m_ee min
\newcommand{\ml}{m_\mathrm{lightest}} % m_ee

\newcommand{\mbbmz}{m_{\beta \beta,0}^{\min}}
\newcommand{\mbbx}{m_{\beta \beta,x_i}}

\newcommand{\NH}{NH} % Normal Hierarchy
\newcommand{\IH}{IH} % Inverted Hierarchy

\newcommand{\el}{\text{e}} % electron

\newcommand{\meV}{\text{meV}}	\newcommand{\eV}{\text{eV}}

\begin{document}

	\title{Empirical inference on the Majorana mass of the ordinary neutrinos}
	\author{Stefano Dell'Oro}
		\email{sdelloro@vt.edu}
		\affiliation{Center for Neutrino Physics, Virginia Polytechnic Institute and State University, Blacksburg, VA 24061, USA \\}%
	\author{Simone Marcocci}
%		\email{mail_tua_o_gssi?}
%		\thanks{Deceased on August 7h, 2019.}
		\affiliation{Fermi National Accelerator Laboratory, Batavia, IL 60510, USA \\}%
	\author{Francesco Vissani}
		\email{francesco.vissani@lngs.infn.it}
		\affiliation{INFN, Laboratori Nazionali del Gran Sasso, 67100 Assergi, L'Aquila, Italy \\}%
		\affiliation{Gran Sasso Science Institute, 67100 L'Aquila, Italy \\}%
		\affiliation{Institute for Advanced Studies (IdEA), University of Campinas, 13083-876 S\~ao Paulo, Brazil \\}%
\date{\today}
	
	\begin{abstract}
		There is a broad theoretical consensus on the idea that ordinary neutrinos have a Majorana mass, but we have no clear prediction about its value, and direct experimental measurements 
		of this quantity are rather challenging.
		In this work, we argue that the current cosmological measurements allow us to obtain precise information on the effective Majorana mass, i.\,e.\ the electronic-type mass of 
		ordinary neutrinos. We show that the numerical results that we obtain can be accurately reproduced, and hence tested, by a straightforward analytical procedure.
		We then discuss the stability of the assumptions at the basis of our analysis and the implications of our findings for neutrinoless double beta decay.
		\\[+9pt]
%		Accepted for publication on \href{https://journals.aps.org/prd/accepted/1907aQ4cUff1832f73e61cb92d5c59bf7feb2cd1b}{Phys. Rev. D}
		Published on: \href{https://journals.aps.org/prd/abstract/10.1103/PhysRevD.100.073003}{Phys.\ Rev.\ D {\bf 100}, 073003 (2019)}
	\end{abstract}

	\maketitle
%------------------------------------------------
\section{Introduction}

	There is compelling evidence from oscillation phenomena~\cite{Fukuda:1998mi,Ahmad:2002jz,Eguchi:2002dm,Michael:2006rx,Agafonova:2015jxn} that ordinary neutrinos are not massless.
	However, we do not have empirical knowledge of the absolute values of these masses. Nor do we know whether these are of Dirac type --- as for all charged fermions --- or of Majorana type.
	
	A Majorana nature for the neutrino mass is considered quite natural from the theoretical point of view~\cite{Minkowski:1977sc,Yanagida:1979as,GellMann:1980vs,Mohapatra:1979ia}.
	This hypothesis could be experimentally tested via the observation of neutrinoless double beta decay
	(\bb), the transition $(A,Z) \to (A,Z+2) + 2 \el^-$ where a nucleus increases its charge by two units, emitting two electrons but no neutrinos~\cite{Furry:1939qr}.
	
	The rate of \bb~depends quadratically upon a combination of the neutrino masses $m_i$, the mixing matrix $U_{\el i}$ and the Majorana phases, which is called the ``effective Majorana mass''
	and, within the three-ordinary-neutrino framework, is defined as
	\begin{equation}
	\label{eq:mbb}
	\mbb \equiv \left| \sum_{i=1,2,3} m_i \left|U_{\el i}^2\right| e^{i\xi_i} \right|.
	\end{equation}
	A key issue is therefore to understand which is the actual value of $\mbb$.
	
	The theory of fermion masses is not yet sufficiently developed to give definitive answers.
	For instance, it has been argued that a plausible value for $\mbb$ could be of the order of $ m \cdot \theta_\mathrm{C}^n$, where $\theta_\mathrm{C} \simeq 13^\circ$ is the Cabibbo 
	mixing angle, $m \equiv  \sqrt{\Delta m^2_\mathrm{atm}} \simeq 50\,\meV$ and $n=1,2$~\cite{Vissani:1998xg,DellOro:2018jze}.
	Still, these indications only offer an ``educated guess'' pointing towards the $\mathcal{O}(1\,\meV)$ scale and might be correct within a factor of a few, 
	which practically means that we do not know $\mbb$ reliably.
	Principled models, on the other hand, such as $SO(10)$ theories, offer more convincing explanations for light Majorana neutrino masses.
	However, these motivations have been extensively investigated~\cite{Matsuda:2001bg,Bajc:2005zf,Bertolini:2009qj,Joshipura:2011nn,Abud:2012xp,Altarelli:2013aqa,Dueck:2013gca,Ohlsson:2019sja},
	but they remain far from being unique and we lack clear criteria on how to assess them.  
	
	In this situation, it seems reasonable to fully exploit the information coming from the experimental observations in order to try to quantify $\mbb$.
	Neutrino oscillations are a powerful tool in this regard, and the mixing angles and mass splittings are currently measured quite precisely~\cite{Capozzi:2018ubv,deSalas:2017kay,Esteban:2018azc}.
	In particular, the interpretation of the oscillation phenomena is beginning to prefer the so-called normal hierarchy (\NH) over the inverted hierarchy (\IH) for the neutrino mass spectrum, 
	now at a level greater than $3\sigma$ (see the Appendix).
	However, neutrino oscillations only measure the differences between the neutrino mass eigenstates $m_i^2-m_j^2$ and not the absolute masses.
	
	The most relevant information on absolute masses to date comes from cosmological measurements.
	It is fair to state that a cautious approach is highly advisable while dealing with the results from cosmological surveys, 
	due to the crucial role of the theoretical assumptions in building the specific model.
	On the other hand, this cannot provide an excuse in order to ignore a whole category of results. 
	Cosmology is able to probe the value of the sum of the active neutrinos that, consistently with the hypothesis of three light neutrinos, is defined as
	$\Sigma \equiv m_1 + m_2 + m_3$.
	Very stringent upper limits $\Sigma < 180\,\meV$ at the $2\sigma$\,C.\,L.\ have been obtained by many independent analyses
	\cite{Ade:2015xua,Palanque-Delabrouille:2015pga,DiValentino:2015sam,Zhang:2015uhk,Cuesta:2015iho,Giusarma:2016phn,Yeche:2017upn,Aghanim:2018eyx,RoyChoudhury:2019hls}. 
	
	\medskip
	In a previous work, we showed the importance of the combination of the results of post-2015 Planck observations and ``small scale'' cosmology measurements for \bb~\cite{Dell'Oro:2015tia}.
	Here, we demonstrate how to obtain detailed information on $\mbb$, possibly measuring this important parameter.
	We first describe a direct Monte Carlo procedure to derive the likelihood of $\mbb$ from that of $\Sigma$ in Sec.~\ref{sec:MC}.
	In Sec.~\ref{sec:analytic}, we test the results we obtain by means of an analytical calculation.
	Finally, in Sec.~\ref{sec:discussion}, we critically discuss the assumptions underlying the present analysis and examine the implications of the predictions we obtain for \bb.
	The main results of this work are summarized in Figs.~\ref{fig:mbb_ml} and \ref{fig:mbb_1D}, that are discussed in Secs.~\ref{sec:MC} and \ref{sec:analytic}.

%------------------------------------------------
\section{Distribution of $\mbb$ from data}
\label{sec:MC}

	%%% FIGURE %%%
	\begin{figure}[tb]
		\centering
		\includegraphics[width=1.\columnwidth]{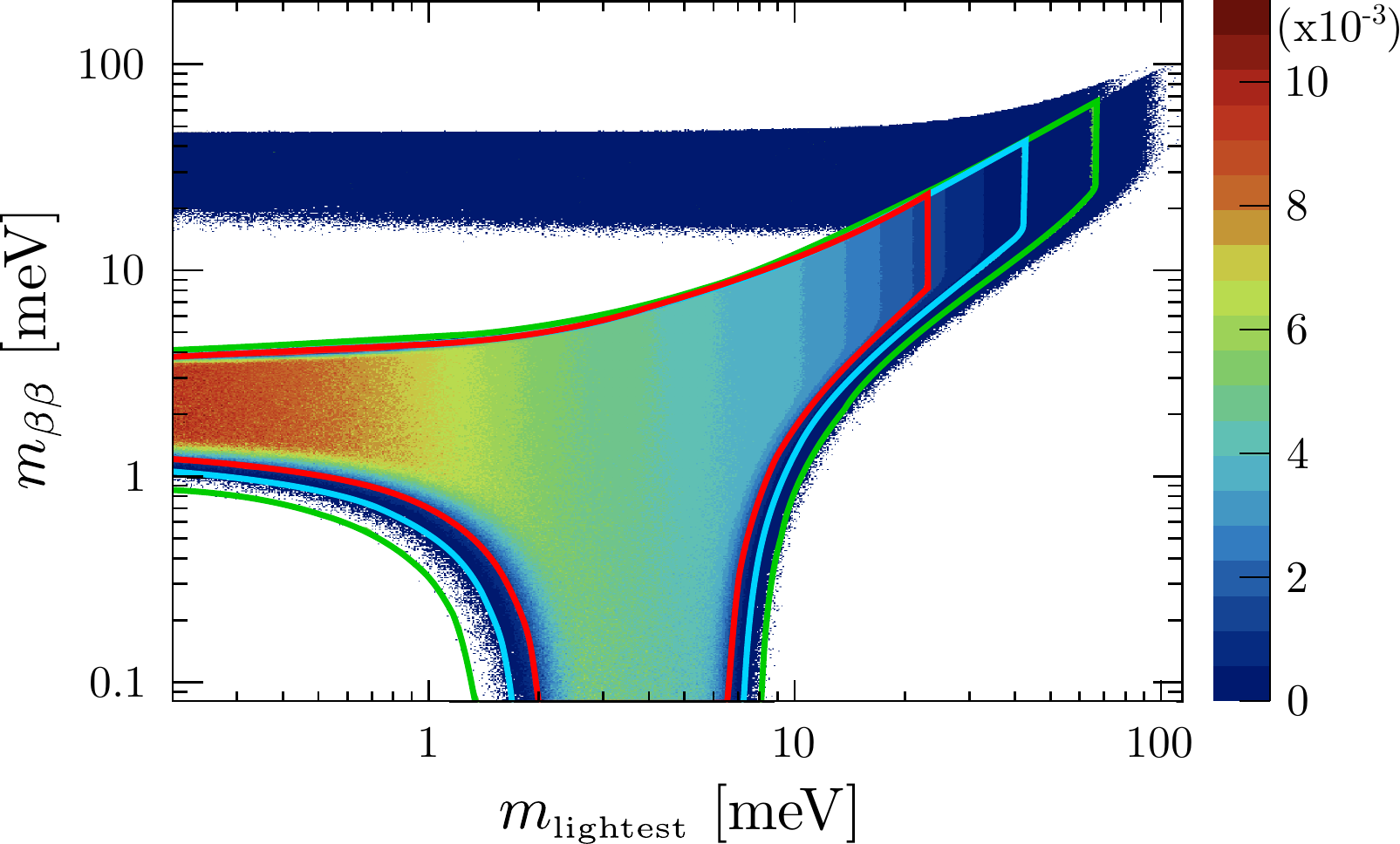}
		\caption{Density plot of $\mbb$ as a function of $\ml$ as a result of the Monte Carlo simulation. The two regions correspond to the \NH~and \IH~cases.
			 For each event, the mass hierarchy is chosen according to the preference evidenced in Ref.~\cite{Capozzi:2018ubv}.
			 The red, blue and green contours are the 1$\sigma$, 2$\sigma$ and 3$\sigma$ C.\,L.\ intervals.}
		\label{fig:mbb_ml}
	\end{figure}

	In a frequentist approach, one would assume each value of $\mbb$ lying below the current experimental bound to be equally allowed.
	In the case of \NH, this would imply the impossibility of excluding that this parameter is actually negligibly small.
	However, this is not necessarily the case. In fact, it is possible to obtain useful information on $\mbb$ independently of that coming from the direct experimental search for \bb.
	
	As an illustrative case, let us suppose that one day cosmology will be able to measure a value of $\Sigma$, and that its value will turn out to be one of the current $2\sigma$ upper limits, 
	i.\,e.\ $140\,\meV$~\cite{Yeche:2017upn}. Since the mass splittings are reliably known, the lightest neutrino mass would be set to $38\,\meV$.
	In turn, still assuming that the neutrinos have a Majorana mass, we could conclude that $\mbb$ lies in the range $(13-39)\,\meV$.
	In particular, the possibility that $\mbb=0$ would  be excluded for experimental reasons.
	
	Passing to a more realistic case, today we can dispose of a limit $\Sigma$ from cosmology, but not a measurement.
	Still, the range of allowed values of $\mbb$ (for a given $\Sigma$) can only be explored by declaring a \emph{prior} distribution for some relevant parameter --- that is, 
	by adopting a Bayesian approach.
	
	A prediction for the possible value of $\mbb$ can by obtained by simulating a large series of combinations of neutrino mass parameters, starting from the information available today.
	We thus developed a Monte Carlo tool that iteratively extracts values of $\mbb$ by providing as input randomly generated values of the mass splittings and mixing angles, and of $\Sigma$.
	The former are picked around the best-fit values of the global analysis of Ref.~\cite{Capozzi:2018ubv} assuming Gaussian errors. 
	The latter parameter is selected within the limit reported in Ref.~\cite{Yeche:2017upn}, in which the authors obtained an almost Gaussian likelihood (see Fig.~10 in the reference)
	that can be approximated by the expression:
	\begin{equation}
	\label{eq:likeCosm}
		 G(\Sigma) = \frac{1}{45.4\,\meV} \, \exp \left[ - \frac{1}{2} \frac{(\Sigma - 41.3\,\meV)^2}{(49.7\,\meV)^2} \right].
	\end{equation}
	Indeed, starting from $G(\Sigma)$ of Eq.~\eqref{eq:likeCosm}, we are able to reproduce the same limit $\Sigma <140\,\meV$ at the $2\sigma$\,C.\,L.\, reported in the reference.
%	\footnote{The Gaussian fit in Eq.~\eqref{eq:likeCosm} actually reproduces the points within $\sim 1\%$ for masses $\gtrsim 50\,\meV$, while it overestimates the likelihood 
%	of a factor up to $\sim 2$ at lower values of $\Sigma$. Anyway, by rejecting the events with negative $\Sigma$ (see the discussion later in the text),
%	we get a limit on $\Sigma$ compatible with that of Ref.~\cite{Yeche:2017upn}.}

	For each event that we simulate, the hierarchy scenario (\NH~or \IH) is initially selected with a preference for \NH, according to the result of the analysis of Ref.~\cite{Capozzi:2018ubv}.
	A value for the lightest neutrino mass, $\ml$, is then calculated starting from $\Sigma$ and the mass splittings~\cite{Dell'Oro:2016dbc}.
	When $\ml$ is negative, i.\,e.\ it falls outside the physical range, the event is rejected and a new one is generated.
	The allowed interval for $\mbb$ is fixed once $\ml$ is known and it is comprised within the extremes~\cite{Vissani:1999tu}
	\begin{subequations}
		\label{eq:mbb_extr}
		\begin{align}
				&\mbbM = \sum_{i=1}^3 \bigl| U_{ei}^2 \bigr| m_i, \label{eq:mbb_max} \\
				&\mbbm = \max \Bigl\{ 2 \bigl| U_{ei}^2 \bigr| m_i - \mbbM, 0 \Bigr\} \quad i=1,2,3 . \label{eq:mbb_min}
		\end{align}
	\end{subequations}

	%%% FIGURE %%%
	\begin{figure}[b]
		\includegraphics[width=1.\columnwidth]{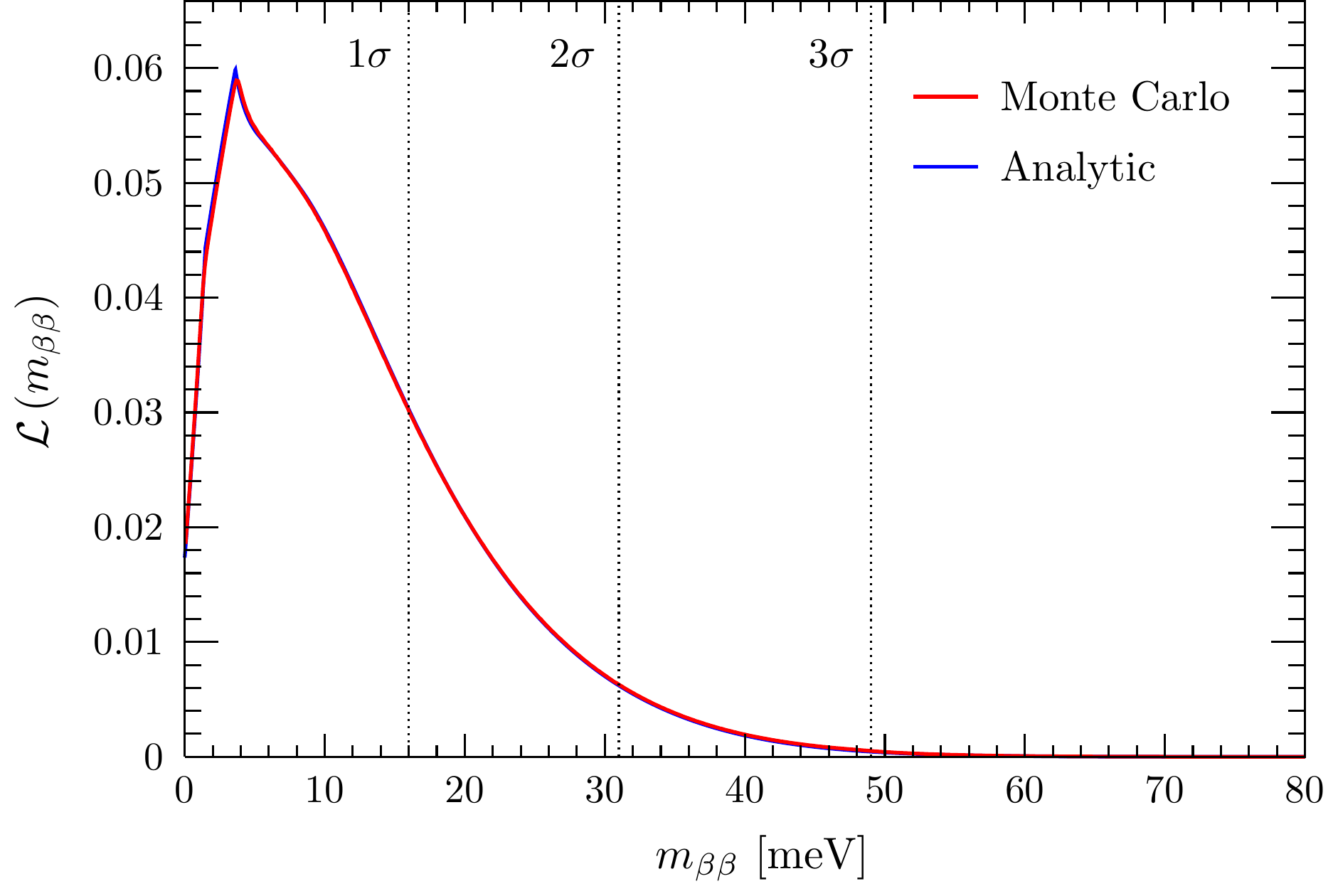}
		\caption{Differential distributions of $\mbb$ obtained by projecting the density plot of Fig.~\ref{fig:mbb_ml} (red line) and as a result of the analytic procedure discussed in the text
			(blue line). The 1$\sigma$, 2$\sigma$ and 3$\sigma$ C.\,L.\ intervals are reported.}
		\label{fig:mbb_1D}
	\end{figure}

\begin{figure}[tb]
		\centering
		\includegraphics[width=1.\columnwidth]{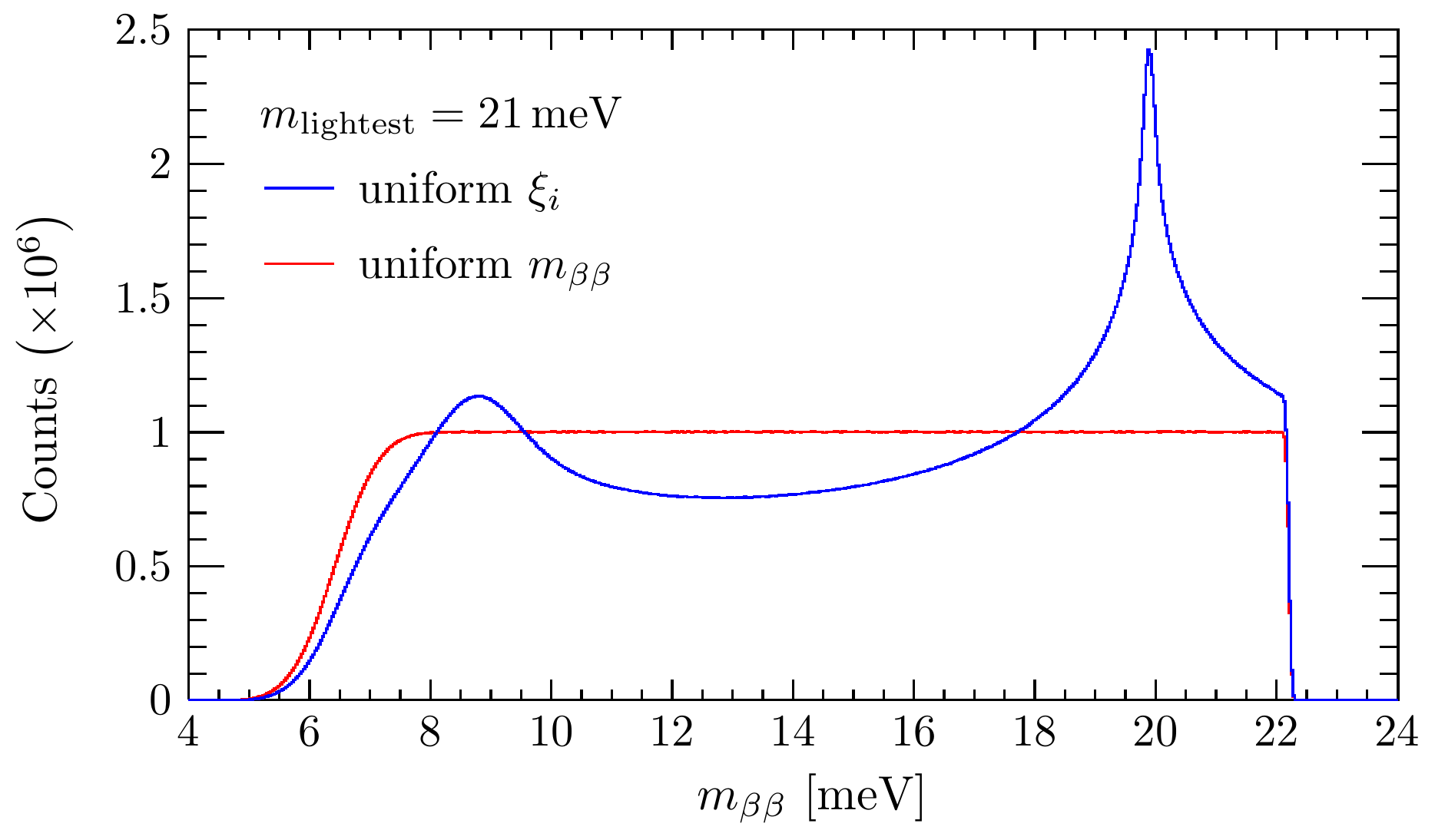}
		\caption{Event distribution for a fixed value of $\ml$ (assuming the \NH~scenario).
			Random mass splittings and mixing angles have been picked around the best-fit values from Ref.~\cite{Capozzi:2018ubv} assuming Gaussian errors, 
			thus fixing the allowed range for $\mbb$.
			Then, two values for $\mbb$ have been extracted assuming a uniform distribution for either $\mbb$ (red) or the Majorana phases (blue).
			The Gaussian shape of the distributions at the boundaries is due to the Gaussian errors on the oscillation parameters.}
		\label{fig:priors}
	\end{figure}

	The value of $\mbb$ is randomly selected within this range with flat probability.
	In principle, other solutions are possible.
	One could adopt the most pessimistic attitude, by setting $\mbb = \mbbm$ or, on the contrary, the most optimistic view, by setting $\mbb = \mbbM$.
	These two extreme choices for the prior of $\mbb$ could be of some interest in order to assess the chances of failure or success in the search for \bb.
	However, neither of them is particularly supported by existing theoretical considerations.
	Also, since our primary goals are to illustrate an analysis methodology and to obtain reasonable expectations on $\mbb$, we did not consider these options.
	A valid alternative, that has already been used in the literature, is to take random Majorana phase uniformly distributed in
	$[0;2\pi)$ and to construct $\mbb$ accordingly~\cite{Benato:2015via,Caldwell:2017mqu,Agostini:2017jim}.
	In this latter case, one typically expects larger values for $\mbb$ up to $\sim 10\%$ since there is a larger probability that the two phases will add constructively, rather than 
	destructively (Fig.~\ref{fig:priors}). This situation is more conducive to the experimental searches.
	However, from a theoretical point of view, it seems easier to motivate a preference for a flat prior on $\mbb$, since this quantity is more directly connected 
	to the Majorana mass term of the Lagrangian density, and thus we opted for this choice. 

	The result of the simulation is shown in Fig.~\ref{fig:mbb_ml}.
	In total, one billion events have been generated.
	The contour lines at different C.\,L.\ are drawn by ``cutting'' the distribution on the $(\ml:\mbb)$ plane at fixed ratios of the fraction of surviving events over the total one.
	The \IH~region is interested by no 1, 2, 3$\sigma$ contour lines since this scenario is excluded at more than $3\sigma$ by the oscillation studies.
	The tight limits on $\Sigma$ push towards smaller values of $\ml$ and this in turn favors smaller values of $\mbb$.
	This appears even more clearly by projecting the plot onto the $\mbb$ axis (Fig.~\ref{fig:mbb_1D}).
	These results will be further discussed in the following sections.
	
	%%% FIGURE %%%
	
%------------------------------------------------
\section{Analytic test of the results}
\label{sec:analytic}

	In order to test our Monte Carlo tool, we developed an analytic procedure that allowed us to obtain a distribution of $\mbb$ to be compared to that derived in the previous section 
	(see in particular Fig.~\ref{fig:mbb_1D}).
	We focused on the \NH~scenario, so that the lightest $m_i$ is now $m_1$ (recall Eq.~\eqref{eq:mbb}).

\begin{figure}[tb]
		\includegraphics[width=0.5\textwidth,keepaspectratio]{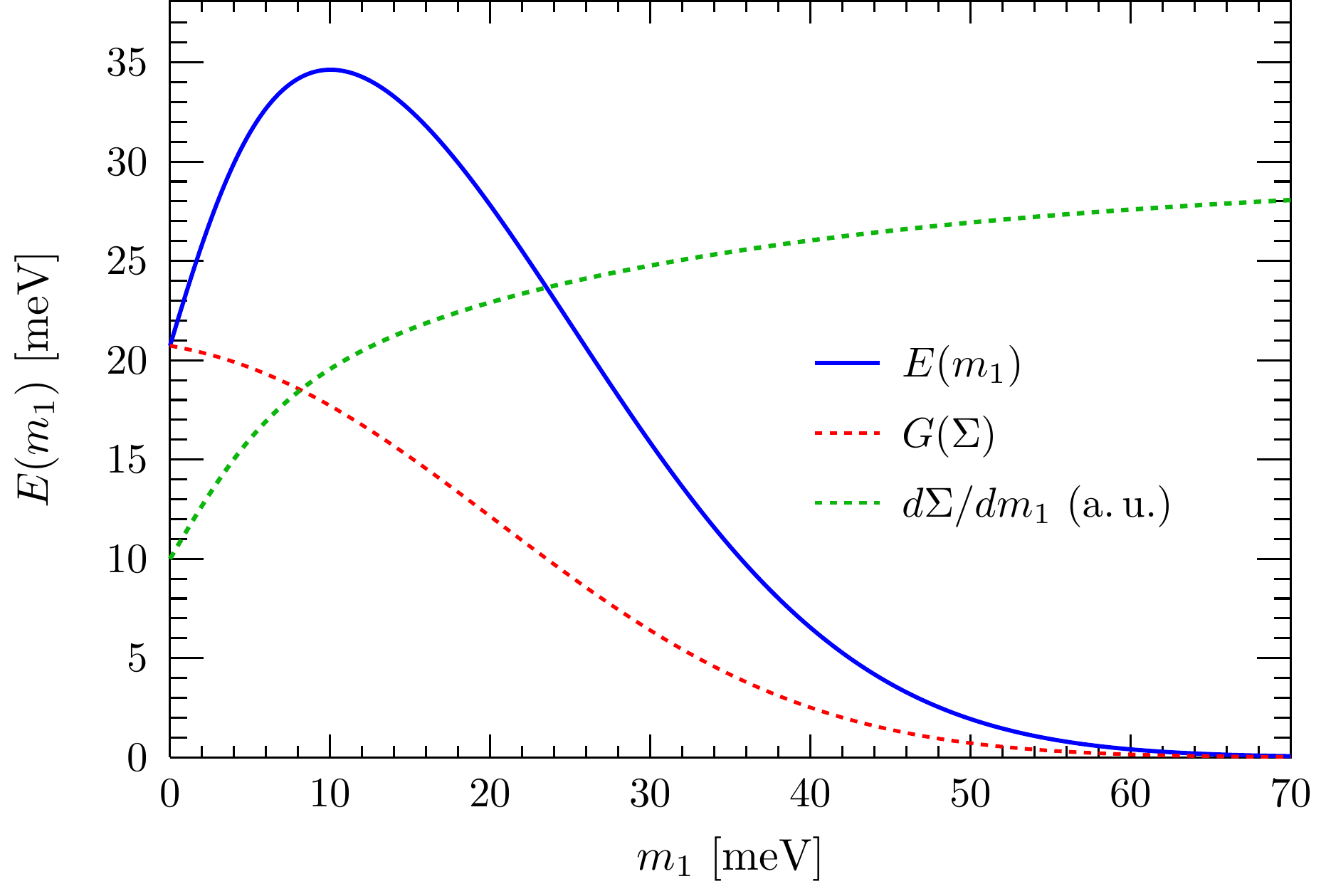}
		\caption{Distribution of the lightest neutrino mass given the information from cosmology $G(\Sigma)$. The decreasing trend at small values of $m_1$ is due to the Jacobian $d\Sigma / dm_1$.}
		\label{fig:distr_m1}
	\end{figure}

	As a starting point, let us consider \emph{a priori} a flat distribution for each fixed $m_1$, namely
	\begin{equation}
		\label{eq:mbb_prior}
		d\mathcal{L}^\mathrm{prior} \left( \mbb \right) = \frac{d \mbb}{\mbbM (m_1) - \mbbm (m_1)}
	\end{equation}
	where the extremes have the known expressions reported in Eqs.~\eqref{eq:mbb_max}--\eqref{eq:mbb_min}.
	As previously stated, we assume a uniform distribution for $\mbb$ within the allowed range.
	The information on the distribution of $m_1$ required by Eq.\eqref{eq:mbb_prior} can be extracted from cosmology.
	The constraints on $\Sigma$ can be summarized by the likelihood
	\begin{equation}
		d\mathcal{L} (\Sigma\,|\,\text{cosm}) = G (\Sigma)\,d\Sigma 
	\end{equation}
	where we can consider $\Sigma$ as a function of $m_1$.
	We use the expression in Eq.~\eqref{eq:likeCosm} and compute the normalization by integrating $G(\Sigma)$ down to $\Sigma = 58.5\,\meV$, corresponding to the case $m_1=0$, 
	given the present values of the mass splittings~\cite{Capozzi:2018ubv}.
	We thus obtain 
	\begin{equation}
		d\mathcal{L} (m_1\,|\,\text{cosm}) = G(\Sigma)\ \frac{d\Sigma}{dm_1} \ dm_1 = E(m_1) \, dm_1.
	\end{equation}
	The distribution of the lightest neutrino mass $E(m_1)$ is shown in Fig.~\ref{fig:distr_m1}.
	The curve has a maximum around $m_1 = 10$\,meV since the Jacobian $d\Sigma/dm_1$, which is a strictly increasing function, disfavors the values close to $m_1 = 0$, while the likelihood 
	extracted from Ref.~\cite{Yeche:2017upn} does the opposite.

	It is worth noticing that the propagation of the uncertainties on the mass splitting values introduces a variation in the distribution $E(m_1)$.
	However these are of the order of a few parts per 10,000.
	Therefore, we can neglect them. %(much smaller than that of cosmology).
	
	The bidimensional distribution of $\mbb$ and $m_1$, which is automatically normalized, is then
	\begin{equation}
	\begin{split}
		d\mathcal{L}(\mbb, m_1) =\  &d\mathcal{L}^\mathrm{prior} (\mbb) \times d\mathcal{L} (m_1\,|\,\text{cosm})  \\[5pt] &= H(m_1) \ d \mbb \ dm_1
	\end{split}
	\end{equation}
	where
	\begin{equation}
		H( m_1) = \frac{E(m_1)}{\mbbM (m_1) - \mbbm (m_1)}.
	\end{equation}
	By integrating over all the possible values of $m_1$, we get the \emph{posterior} distribution of $\mbb$, which includes the information extracted from cosmology:
	\begin{equation}
	\label{eq:L_diff}
	\begin{split}
		d\mathcal L^\mathrm{posterior} &(\mbb) \equiv d\mathcal L (\mbb\,|\,\text{cosm}) \\[5pt] &= F(\mbb) \ d \mbb
	\end{split}
	\end{equation}
	where
	\begin{equation}
		F(\mbb)  = \int_{m_1^{\min} (\mbb)} ^{m_1^{\max} (\mbb)} H(m_1) \ dm_1.
	\end{equation}

	%%% FIGURE %%%

	The functions describing the extreme values of $m_1$ given $\mbb$ can be analytically computed, since they correspond to solvable fourth-degree equations.
	The allowed region in the $(m_1: \mbb)$ plane is that shown in Fig.~\ref{fig:mbb_ml}, keeping in mind that we are now considering $\mbb$ as the independent variable.

	The double integral we need to solve in order to get the probability of a specific $\mbb^*$, i.\,e.\ the cumulant distribution $\mathcal L (\mbb^*) \equiv \mathcal L (\mbb < \mbb^*)$,
	can actually be rearranged into a one-dimensional one:
	\begin{align}
%		\mathcal L (\mbb^*) = \int_{\text{lower} (\mbb^*)} ^{\text{upper} (\mbb^*) } \\ \frac{\min \left[ \mbbM (m_1), \mbb^* \right] - \mbbm (m_1)}{\mbbM (m_1) - \mbbm (m_1)}\,E( m_1)\,dm_1
		\mathcal L (\mbb^*) = &\int_{\text{lower} (\mbb^*)} ^{\text{upper} (\mbb^*) } E( m_1)\,dm_1\\ 
		&\cdot \frac{\min \left[ \mbbM (m_1), \mbb^* \right] - \mbbm (m_1)}{\mbbM (m_1) - \mbbm (m_1)}. \notag
	\end{align}
	The upper limit of integration is
	\begin{equation}
		\text{upper}(\mbb^*) = m_1^{\max} (\mbb^*)
	\end{equation}
	while we can identify two cases for the inferior one, depending on the value of $\mbb^*$ with respect to $\mbbmz \equiv \mbbm (m_1=0)$, namely
	\begin{equation}
	\label{eq:lower}
		\text{lower} (\mbb^*) = \left\{
		\begin{array}{ll}
			m_1^{\min} (\mbb^*)	&\mbox{ if } \mbb^* < \mbbmz  \\[5pt] 
			0							&\mbox{ if } \mbb^* > \mbbmz
		\end{array}
		\right. .
	\end{equation}

	If $\mbb^* \to 0$, then $\min \left[ \mbbM (m_1), \mbb^* \right] = \mbb^*$ and the numerator $\to 0$. Instead, if $\mbb^* \to \infty$, then 
	$\min \left[ \mbbM (m_1), \mbb^* \right]  = \mbbM (m_1)$, the fraction approaches $1$, the inferior integration limit becomes $0$,
	and the integral of $E(m_1)$ over all values is indeed $1$.
	In practice, for sufficiently high confidence levels (corresponding to larger values of $\mbb^*$), we will refer to the latter value of Eq.~\eqref{eq:lower}.
	Therefore, we can rewrite
	\begin{align}
	\label{eq:like_post}
		\mathcal L &(\mbb^*) = \int_{0} ^{m_1^{\min} (\mbb^*)} E(m_1)\,dm_1 \\[5pt]
		&+ \int_{m_1^{\min} (\mbb^*)} ^{m_1^{\max} (\mbb^*)} \frac{\mbb^* - \mbbm (m_1)}{\mbbM (m_1) - \mbbm (m_1)}\,E( m_1)\,dm_1. \notag
	\end{align}
	By using Eq.~\eqref{eq:like_post}, we can compute $\mbb$ for each given confidence level by setting $\mathcal{L} (\mbb^*) = $ C.\,L.\ and choosing the oscillation parameters at their central 
	(best-fit) values.

	The value of $\mbb$ shifts when we include the variation on the mixing angles and mass splittings, since these enter both the integration limits and the integrand of Eq.~\eqref{eq:like_post}.
	The individual contribution associated to the uncertainty of each oscillation parameter $x_i$ can be analytically computed. For a fixed confidence level, we get:
	\begin{equation}
%		\Delta \mbb ^* = \sqrt{ \sum_{x_i = \mathrm{\,osc.\ par.}} \left(  \Delta x_i \, \frac{ \partial \mathcal L (\mbb^*)}{\partial x_i}  \right)^2} \,
%		\left| \frac{\partial \mathcal L (\mbb^*)}{\partial \mbb^*} \right|^{-1}
		\Delta \mbbx ^* = \Delta x_i \, \frac{ \partial \mathcal L (\mbb^*)}{\partial x_i}\
		\left| \frac{\partial \mathcal L (\mbb^*)}{\partial \mbb^*} \right|^{-1}
	\end{equation}
	evaluated in the best-fit central region. 
	In practice, the overall effect is numerically small and the impact is negligible for our purposes.

	In order check the consistency between the two described procedures, we can compare the resulting differential distributions for $\mbb$.
	The distribution from the likelihood of Eq.~\eqref{eq:L_diff} can directly be compared with that obtained by the Monte Carlo simulation.
	The result, illustrated in Fig.~\ref{fig:mbb_1D}, shows a very good agreement between the two procedures.
	This is a rather interesting conclusion: on the one hand, we have validated the numerical results. On the other hand, we have learned that the analytical method described here 
	produces accurate results. 
  
%------------------------------------------------
\section{Discussion}
\label{sec:discussion}
	
	The sum of the active neutrino mass coming from cosmology is a key ingredient of the analysis we presented.
	Therefore, we would like to begin our discussion with some considerations on the robustness of the current measurement of $\Sigma$.
	
	To this extent, a first issue to address is the possible existence of additional light neutrinos, beyond those envisaged by the Standard Model. 
	The presence of new neutral fermions was proposed a long time ago~\cite{Pontecorvo:1957qd}. 
	Despite the fact that the values of their masses and mixing with the known neutrinos are \emph{a priori} unknown, the interpretation of some experiments, when individually taken, 
	would be compatible with the existence of neutrinos with mass of $\mathcal{O}(1\,\eV)$.
	These neutrinos would give rise to observable oscillation effects.
	On the other hand, the various analyses that have extensively considered the implications of this hypothesis highlighted the difficulties that arise,
	and even the inconsistencies, when compared to the available data~\cite{Cirelli:2004cz,Dentler:2018sju}.
	Therefore, while this could lead to more complicated scenarios for these new neutrinos~\cite{Diaz:2019fwt},
	the ``minimal'' hypothesis of three ordinary neutrinos considered here is not weakened.%
	\footnote{Interestingly, $1\,\eV$-neutrinos could have a significant impact on the rate of \bb~\cite{Girardi:2013zra,Giunti:2015kza,Huang:2019qvq}.}

	Recently, new results on the Hubble constant in the local Universe motivated the idea of rethinking the accepted cosmological model~\cite{Riess:2019cxk}.
	In principle, this could have an impact on the conclusions of our work.
	The new observations would be explained by the presence of some ``kind'' of sterile neutrinos.
	These neutrinos should have some specific proprieties, namely they should self-interact via strong interactions~\cite{Verde:2019ivm}
	and should have rather small masses, subject to the stringent limits of cosmology~\cite{Aghanim:2018eyx}.
	In other words, while the assumption that there are three active neutrinos in cosmology could be reconsidered, we do not have any evidence that 
	the measurement of $\Sigma$ should be affected.
	These considerations can be quantified by an example.
	Such a type of neutrino, with mass $m_0$ slightly larger than $m_1$, had already been introduced, not only to increase the number of effective neutrinos in cosmology,
	but also to explain certain features of the observed solar neutrino spectrum~\cite{deHolanda:2010am}.
	The resulting mixing $ U_{\el0}^2\sim 10^{-3}$ and mass splitting $m_0^2-m_1^2\sim 0.2\ \Delta m^2_\odot$~\cite{deHolanda:2010am}
	would add a contribution $U_{\el 0}^2 (m_0-m_1)$ to $\mbb$ of the order of $10^{-3}$\,meV at most, and therefore it is completely negligible. 

	In summary, it seems premature to us to conclude that the the minimal set of assumptions adopted here and the reported results should require significant corrections.
  
	To conclude our discussion, we would like to consider the implications of our findings for \bb.
	Possible predictions for the expected value of $\mbb$ could be of great interest for the experimental community of \bb~in helping to understand the chances for a positive observation,
	or to figure out how far from reach a promising target might be.
	
	The current \bb~experiments have already set limits of the order of $(10^{25}-10^{26})$\,yr on the decay 
	half-life~\cite{Alvis:2019sil,Agostini:2018tnm,Alduino:2017ehq,Anton:2019wmi,KamLAND-Zen:2016pfg}.
	The corresponding upper bounds for $\mbb$ are around $100$\,meV, with values that actually span a large range, mostly due to the theoretical uncertainties coming from nuclear 
	physics~\cite{Dell'Oro:2014yca}, which is a fundamental ingredient in order to extract the information on the neutrino mass.
	At the same time, the forthcoming generation of experiments is setting very ambitious goals, aiming at sensitivities greater than $10^{27}$\,yr.
	This should allow to explore the parameter space of $\mbb$ of the order of tens of meV~\cite{Agostini:2017jim}.
	
	Given the present and near-future experiment sensitivities, we can thus analyze the results shown in Fig.~\ref{fig:mbb_1D}.
	The differential distribution of $\mbb$ is peaked at $4\,\meV$, while the $1\sigma$, $2\sigma$ and $3\sigma$ intervals extends up to 16, 31 and 49\,meV, respectively.
	On the one hand, this means that we should be able to begin to probe the parameter region of interest in the coming years.
	On the other hand, this means that we are still quite far from fully exploring the ``core'' of the distribution, i.\,e.\ values of $\mbb$ of the order of a few meV.
	This ultimate investigation would require an extremely challenging multi-tonne experiment~\cite{Biller:2013wua}, but this hypothesis should not be ruled out.

	The search for \bb~remains one of the main ways, if not the only one, to address some major open questions in particle physics: the conservation of the lepton number conservation, 
	and the value of the neutrino mass and its nature!

%------------------------------------------------
\section{Summary}

	We developed a Monte Carlo tool to extract the information on the effective Majorana mass starting from the available data on the neutrino masses, coming from the
	oscillation studies and from cosmology.
	We found that the distribution of $\mbb$ tends toward low values of the parameter region, with a mode at $4\,\meV$, and a $3\sigma$ interval extended up to almost 50\,meV.
	We validated our results with an analytical procedure, whose outcome perfectly matches the numerical one.
	Finally, we discussed the assumptions at the basis of our analysis and implications of the new information for neutrinoless double beta decay.
	
	\begin{acknowledgments}
		This work was partially supported by the research grant number 2017W4HA7S ``NAT-NET: Neutrino and Astroparticle Theory Network'' 
		under the program PRIN 2017 funded by the Italian Ministero dell'Istruzione, dell'Universit\`a e della Ricerca (MIUR). 
	
		\medskip
		This work is dedicated to the memory of Simone, a very dear friend and a brilliant scientist, who passed away on August 7th, 2019.
		He helped in the fulfillment of this article until the very end.
		We will deeply miss him.
	\end{acknowledgments}

\appendix*
%------------------------------------------------
\section{On the preference for the Normal Hierarchy}
	\label{app:NH_preference}
	
	In the global analyses, it is customary to present the preference for the hierarchy scenario in terms of the difference 
	$\Delta \chi^2 \equiv \chi^2_\mathrm{IH} - \chi^2_\mathrm{NH} >0$ between the two cases.
	This quantity has a different meaning than in the case of one-dimensional parameters.
%	and it needs to be interpreted accounting for the fact that it concerns a discrete degree of freedom.
	In order to determine how much is \NH~preferred by the data, we can estimate the likelihood
	\begin{equation}
	\begin{split}
		\mathcal{L}(\mbox{IH}) &= \frac{\mathcal{L}^\mathrm{cosm}_\mathrm{IH} \times \el^{-\chi^2_\mathrm{IH}}}%
			{\mathcal{L}^\mathrm{cosm}_\mathrm{IH} \times \el^{-\chi^2_\mathrm{IH}} + \mathcal{L}^\mathrm{cosm}_\mathrm{NH} \times \el^{-\chi^2_\mathrm{NH}}} \\
		&= \left( 1 + \frac{\mathcal{L}^\mathrm{cosm}_\mathrm{NH} \times \exp( {\Delta \chi^2/2} )}{1 - \mathcal{L}^\mathrm{cosm}_\mathrm{NH}} \right)^{-1}
	\end{split}
	\end{equation}
	which combines the independent information from cosmology, namely $\mathcal{L}^\mathrm{cosm}_\mathrm{NH}$
	($\mathcal{L}^\mathrm{cosm}_\mathrm{IH} \equiv 1 - \mathcal{L}^\mathrm{cosm}_\mathrm{NH}$),
	and that from neutrino oscillations.
	The analytical procedure described in the text or, equivalently, the Monte Carlo tool, gives $\mathcal{L}^\mathrm{cosm}_\mathrm{NH} \simeq 0.75$.
	In other words, the \NH~is about 3 times more probable than the \IH.
	
	The slightly different values of $\Delta \chi^2 = \{ 9.5,\ 11.7,\  9.3 \}$ were reported in Refs.~\cite{Capozzi:2018ubv}, \cite{deSalas:2017kay} and \cite{Esteban:2018azc}, respectively.
	These values give $\mathcal{L}(\mbox{IH}) =\{ 2.9 ,\ 1.0,\ 3.2 \}\cdot 10^{-3}$, which can be presented in the language of Gaussian distributions as $\{ 3.0,\ 3.3,\ 3.0 \} \sigma$. 
	If the information from cosmology was omitted, setting arbitrarily $\mathcal{L}^\mathrm{cosm}_\mathrm{NH} = 0.5$, 
	the same results would reduce to $\{ 2.6,\ 3.0,\ 2.6 \} \sigma$.

%------------------------------------------------
	\bibliography{ref_arXiv,ref_Papers,ref_Proceedings}

%merlin.mbs apsrev4-1.bst 2010-07-25 4.21a (PWD, AO, DPC) hacked
%Control: key (0)
%Control: author (8) initials jnrlst
%Control: editor formatted (1) identically to author
%Control: production of article title (-1) disabled
%Control: page (0) single
%Control: year (1) truncated
%Control: production of eprint (0) enabled
\begin{thebibliography}{55}%
\makeatletter
\providecommand \@ifxundefined [1]{%
 \@ifx{#1\undefined}
}%
\providecommand \@ifnum [1]{%
 \ifnum #1\expandafter \@firstoftwo
 \else \expandafter \@secondoftwo
 \fi
}%
\providecommand \@ifx [1]{%
 \ifx #1\expandafter \@firstoftwo
 \else \expandafter \@secondoftwo
 \fi
}%
\providecommand \natexlab [1]{#1}%
\providecommand \enquote  [1]{``#1''}%
\providecommand \bibnamefont  [1]{#1}%
\providecommand \bibfnamefont [1]{#1}%
\providecommand \citenamefont [1]{#1}%
\providecommand \href@noop [0]{\@secondoftwo}%
\providecommand \href [0]{\begingroup \@sanitize@url \@href}%
\providecommand \@href[1]{\@@startlink{#1}\@@href}%
\providecommand \@@href[1]{\endgroup#1\@@endlink}%
\providecommand \@sanitize@url [0]{\catcode `\\12\catcode `\$12\catcode
  `\&12\catcode `\#12\catcode `\^12\catcode `\_12\catcode `\%12\relax}%
\providecommand \@@startlink[1]{}%
\providecommand \@@endlink[0]{}%
\providecommand \url  [0]{\begingroup\@sanitize@url \@url }%
\providecommand \@url [1]{\endgroup\@href {#1}{\urlprefix }}%
\providecommand \urlprefix  [0]{URL }%
\providecommand \Eprint [0]{\href }%
\providecommand \doibase [0]{http://dx.doi.org/}%
\providecommand \selectlanguage [0]{\@gobble}%
\providecommand \bibinfo  [0]{\@secondoftwo}%
\providecommand \bibfield  [0]{\@secondoftwo}%
\providecommand \translation [1]{[#1]}%
\providecommand \BibitemOpen [0]{}%
\providecommand \bibitemStop [0]{}%
\providecommand \bibitemNoStop [0]{.\EOS\space}%
\providecommand \EOS [0]{\spacefactor3000\relax}%
\providecommand \BibitemShut  [1]{\csname bibitem#1\endcsname}%
\let\auto@bib@innerbib\@empty
%</preamble>
\bibitem [{\citenamefont {Fukuda}\ \emph {et~al.}(1998)\citenamefont {Fukuda}
  \emph {et~al.}}]{Fukuda:1998mi}%
  \BibitemOpen
  \bibfield  {author} {\bibinfo {author} {\bibfnamefont {Y.}~\bibnamefont
  {Fukuda}} \emph {et~al.} (\bibinfo {collaboration} {Super-Kamiokande
  Collaboration}),\ }\href {\doibase 10.1103/PhysRevLett.81.1562} {\bibfield
  {journal} {\bibinfo  {journal} {Phys.\ Rev.\ Lett.}\ }\textbf {\bibinfo
  {volume} {81}},\ \bibinfo {pages} {1562} (\bibinfo {year}
  {1998})}\BibitemShut {NoStop}%
\bibitem [{\citenamefont {Ahmad}\ \emph {et~al.}(2002)\citenamefont {Ahmad}
  \emph {et~al.}}]{Ahmad:2002jz}%
  \BibitemOpen
  \bibfield  {author} {\bibinfo {author} {\bibfnamefont {Q.~R.}\ \bibnamefont
  {Ahmad}} \emph {et~al.} (\bibinfo {collaboration} {SNO Collaboration}),\
  }\href {\doibase 10.1103/PhysRevLett.89.011301} {\bibfield  {journal}
  {\bibinfo  {journal} {Phys.\ Rev.\ Lett.}\ }\textbf {\bibinfo {volume}
  {89}},\ \bibinfo {pages} {011301} (\bibinfo {year} {2002})}\BibitemShut
  {NoStop}%
\bibitem [{\citenamefont {Eguchi}\ \emph {et~al.}(2003)\citenamefont {Eguchi}
  \emph {et~al.}}]{Eguchi:2002dm}%
  \BibitemOpen
  \bibfield  {author} {\bibinfo {author} {\bibfnamefont {K.}~\bibnamefont
  {Eguchi}} \emph {et~al.} (\bibinfo {collaboration} {KamLAND Collaboration}),\
  }\href {\doibase 10.1103/PhysRevLett.90.021802} {\bibfield  {journal}
  {\bibinfo  {journal} {Phys.\ Rev.\ Lett.}\ }\textbf {\bibinfo {volume}
  {90}},\ \bibinfo {pages} {021802} (\bibinfo {year} {2003})}\BibitemShut
  {NoStop}%
\bibitem [{\citenamefont {Michael}\ \emph {et~al.}(2006)\citenamefont {Michael}
  \emph {et~al.}}]{Michael:2006rx}%
  \BibitemOpen
  \bibfield  {author} {\bibinfo {author} {\bibfnamefont {D.~G.}\ \bibnamefont
  {Michael}} \emph {et~al.} (\bibinfo {collaboration} {MINOS Collaboration}),\
  }\href {\doibase 10.1103/PhysRevLett.97.191801} {\bibfield  {journal}
  {\bibinfo  {journal} {Phys.\ Rev.\ Lett.}\ }\textbf {\bibinfo {volume}
  {97}},\ \bibinfo {pages} {191801} (\bibinfo {year} {2006})}\BibitemShut
  {NoStop}%
\bibitem [{\citenamefont {Agafonova}\ \emph {et~al.}(2015)\citenamefont
  {Agafonova} \emph {et~al.}}]{Agafonova:2015jxn}%
  \BibitemOpen
  \bibfield  {author} {\bibinfo {author} {\bibfnamefont {N.}~\bibnamefont
  {Agafonova}} \emph {et~al.} (\bibinfo {collaboration} {OPERA
  Collaboration}),\ }\href {\doibase 10.1103/PhysRevLett.115.121802} {\bibfield
   {journal} {\bibinfo  {journal} {Phys.\ Rev.\ Lett.}\ }\textbf {\bibinfo
  {volume} {115}},\ \bibinfo {pages} {121802} (\bibinfo {year}
  {2015})}\BibitemShut {NoStop}%
\bibitem [{\citenamefont {Minkowski}(1977)}]{Minkowski:1977sc}%
  \BibitemOpen
  \bibfield  {author} {\bibinfo {author} {\bibfnamefont {P.}~\bibnamefont
  {Minkowski}},\ }\href {\doibase 10.1016/0370-2693(77)90435-X} {\bibfield
  {journal} {\bibinfo  {journal} {Phys.\ Lett.\ B}\ }\textbf {\bibinfo {volume}
  {67}},\ \bibinfo {pages} {421} (\bibinfo {year} {1977})}\BibitemShut
  {NoStop}%
%%CITATION = PHLTA,B67,421;%%
\bibitem [{\citenamefont {Yanagida}(1979)}]{Yanagida:1979as}%
  \BibitemOpen
  \bibfield  {author} {\bibinfo {author} {\bibfnamefont {T.}~\bibnamefont
  {Yanagida}},\ }\href@noop {} {\bibfield  {journal} {\bibinfo  {journal}
  {Proc.\ Workshop Baryon Number of the Universe and Unified Theories, Tsukuba,
  Japan, February 1979}\ ,\ \bibinfo {pages} {95}} (\bibinfo {year}
  {1979})}\BibitemShut {NoStop}%
%%CITATION = CONFP,C7902131,95;%%
\bibitem [{\citenamefont {Gell-Mann}\ \emph {et~al.}(1979)\citenamefont
  {Gell-Mann}, \citenamefont {Ramond},\ and\ \citenamefont
  {Slansky}}]{GellMann:1980vs}%
  \BibitemOpen
  \bibfield  {author} {\bibinfo {author} {\bibfnamefont {M.}~\bibnamefont
  {Gell-Mann}}, \bibinfo {author} {\bibfnamefont {P.}~\bibnamefont {Ramond}}, \
  and\ \bibinfo {author} {\bibfnamefont {R.}~\bibnamefont {Slansky}},\
  }\href@noop {} {\bibfield  {journal} {\bibinfo  {journal} {Proceedings of the
  Supergravity Workshop, Stony Brook, New York, USA, September 1979}\ }\textbf
  {\bibinfo {volume} {\!\!}},\ \bibinfo {pages} {315} (\bibinfo {year}
  {1979})}\BibitemShut {NoStop}%
%%CITATION = ARXIV:1306.4669;%%
\bibitem [{\citenamefont {Mohapatra}\ and\ \citenamefont
  {Senjanovic}(1980)}]{Mohapatra:1979ia}%
  \BibitemOpen
  \bibfield  {author} {\bibinfo {author} {\bibfnamefont {R.~N.}\ \bibnamefont
  {Mohapatra}}\ and\ \bibinfo {author} {\bibfnamefont {G.}~\bibnamefont
  {Senjanovic}},\ }\href {\doibase 10.1103/PhysRevLett.44.912} {\bibfield
  {journal} {\bibinfo  {journal} {Phys.\ Rev.\ Lett.}\ }\textbf {\bibinfo
  {volume} {44}},\ \bibinfo {pages} {912} (\bibinfo {year} {1980})}\BibitemShut
  {NoStop}%
%%CITATION = PRLTA,44,912;%%
\bibitem [{\citenamefont {Furry}(1939)}]{Furry:1939qr}%
  \BibitemOpen
  \bibfield  {author} {\bibinfo {author} {\bibfnamefont {W.~H.}\ \bibnamefont
  {Furry}},\ }\href {\doibase 10.1103/PhysRev.56.1184} {\bibfield  {journal}
  {\bibinfo  {journal} {Phys.\ Rev.}\ }\textbf {\bibinfo {volume} {56}},\
  \bibinfo {pages} {1184} (\bibinfo {year} {1939})}\BibitemShut {NoStop}%
%%CITATION = PHRVA,56,1184;%%
\bibitem [{\citenamefont {Vissani}(1998)}]{Vissani:1998xg}%
  \BibitemOpen
  \bibfield  {author} {\bibinfo {author} {\bibfnamefont {F.}~\bibnamefont
  {Vissani}},\ }\href {\doibase 10.1088/1126-6708/1998/11/025} {\bibfield
  {journal} {\bibinfo  {journal} {J.\ High Energy Phys.}\ }\textbf {\bibinfo
  {volume} {11}},\ \bibinfo {pages} {025} (\bibinfo {year} {1998})}\BibitemShut
  {NoStop}%
\bibitem [{\citenamefont {Dell'Oro}\ \emph {et~al.}(2018)\citenamefont
  {Dell'Oro}, \citenamefont {Marcocci},\ and\ \citenamefont
  {Vissani}}]{DellOro:2018jze}%
  \BibitemOpen
  \bibfield  {author} {\bibinfo {author} {\bibfnamefont {S.}~\bibnamefont
  {Dell'Oro}}, \bibinfo {author} {\bibfnamefont {S.}~\bibnamefont {Marcocci}},
  \ and\ \bibinfo {author} {\bibfnamefont {F.}~\bibnamefont {Vissani}},\ }\href
  {\doibase 10.1088/1742-6596/1056/1/012059} {\bibfield  {journal} {\bibinfo
  {journal} {J.\ Phys.\ Conf.\ Ser.}\ }\textbf {\bibinfo {volume} {1056}},\
  \bibinfo {pages} {012059} (\bibinfo {year} {2018})}\BibitemShut {NoStop}%
\bibitem [{\citenamefont {Matsuda}\ \emph {et~al.}(2002)\citenamefont
  {Matsuda}, \citenamefont {Koide}, \citenamefont {Fukuyama},\ and\
  \citenamefont {Nishiura}}]{Matsuda:2001bg}%
  \BibitemOpen
  \bibfield  {author} {\bibinfo {author} {\bibfnamefont {K.}~\bibnamefont
  {Matsuda}}, \bibinfo {author} {\bibfnamefont {Y.}~\bibnamefont {Koide}},
  \bibinfo {author} {\bibfnamefont {T.}~\bibnamefont {Fukuyama}}, \ and\
  \bibinfo {author} {\bibfnamefont {H.}~\bibnamefont {Nishiura}},\ }\href
  {\doibase 10.1103/PhysRevD.65.033008, 10.1103/PhysRevD.65.079904} {\bibfield
  {journal} {\bibinfo  {journal} {Phys.\ Rev.\ D}\ }\textbf {\bibinfo {volume}
  {65}},\ \bibinfo {pages} {033008} (\bibinfo {year} {2002})},\ \bibinfo {note}
  {[Erratum:
  \href{https://journals.aps.org/prd/abstract/10.1103/PhysRevD.65.079904}{Phys.\
  Rev.\ D {\bf 65}, 079904 (2002)}]}\BibitemShut {NoStop}%
\bibitem [{\citenamefont {Bajc}\ \emph {et~al.}(2006)\citenamefont {Bajc},
  \citenamefont {Melfo}, \citenamefont {Senjanovic},\ and\ \citenamefont
  {Vissani}}]{Bajc:2005zf}%
  \BibitemOpen
  \bibfield  {author} {\bibinfo {author} {\bibfnamefont {B.}~\bibnamefont
  {Bajc}}, \bibinfo {author} {\bibfnamefont {A.}~\bibnamefont {Melfo}},
  \bibinfo {author} {\bibfnamefont {G.}~\bibnamefont {Senjanovic}}, \ and\
  \bibinfo {author} {\bibfnamefont {F.}~\bibnamefont {Vissani}},\ }\href
  {\doibase 10.1103/PhysRevD.73.055001} {\bibfield  {journal} {\bibinfo
  {journal} {Phys.\ Rev.\ D}\ }\textbf {\bibinfo {volume} {73}},\ \bibinfo
  {pages} {055001} (\bibinfo {year} {2006})}\BibitemShut {NoStop}%
\bibitem [{\citenamefont {Bertolini}\ \emph {et~al.}(2009)\citenamefont
  {Bertolini}, \citenamefont {Di~Luzio},\ and\ \citenamefont
  {Malinsky}}]{Bertolini:2009qj}%
  \BibitemOpen
  \bibfield  {author} {\bibinfo {author} {\bibfnamefont {S.}~\bibnamefont
  {Bertolini}}, \bibinfo {author} {\bibfnamefont {L.}~\bibnamefont {Di~Luzio}},
  \ and\ \bibinfo {author} {\bibfnamefont {M.}~\bibnamefont {Malinsky}},\
  }\href {\doibase 10.1103/PhysRevD.80.015013} {\bibfield  {journal} {\bibinfo
  {journal} {Phys.\ Rev.\ D}\ }\textbf {\bibinfo {volume} {80}},\ \bibinfo
  {pages} {015013} (\bibinfo {year} {2009})}\BibitemShut {NoStop}%
\bibitem [{\citenamefont {Joshipura}\ and\ \citenamefont
  {Patel}(2011)}]{Joshipura:2011nn}%
  \BibitemOpen
  \bibfield  {author} {\bibinfo {author} {\bibfnamefont {A.~S.}\ \bibnamefont
  {Joshipura}}\ and\ \bibinfo {author} {\bibfnamefont {K.~M.}\ \bibnamefont
  {Patel}},\ }\href {\doibase 10.1103/PhysRevD.83.095002} {\bibfield  {journal}
  {\bibinfo  {journal} {Phys.\ Rev.\ D}\ }\textbf {\bibinfo {volume} {83}},\
  \bibinfo {pages} {095002} (\bibinfo {year} {2011})}\BibitemShut {NoStop}%
\bibitem [{\citenamefont {Abud}\ \emph {et~al.}(2012)\citenamefont {Abud},
  \citenamefont {Buccella},\ and\ \citenamefont {Falcone}}]{Abud:2012xp}%
  \BibitemOpen
  \bibfield  {author} {\bibinfo {author} {\bibfnamefont {M.}~\bibnamefont
  {Abud}}, \bibinfo {author} {\bibfnamefont {F.}~\bibnamefont {Buccella}}, \
  and\ \bibinfo {author} {\bibfnamefont {D.}~\bibnamefont {Falcone}},\ }\href
  {\doibase 10.1103/PhysRevD.86.073014} {\bibfield  {journal} {\bibinfo
  {journal} {Phys.\ Rev.\ D}\ }\textbf {\bibinfo {volume} {86}},\ \bibinfo
  {pages} {073014} (\bibinfo {year} {2012})}\BibitemShut {NoStop}%
\bibitem [{\citenamefont {Altarelli}\ and\ \citenamefont
  {Meloni}(2013)}]{Altarelli:2013aqa}%
  \BibitemOpen
  \bibfield  {author} {\bibinfo {author} {\bibfnamefont {G.}~\bibnamefont
  {Altarelli}}\ and\ \bibinfo {author} {\bibfnamefont {D.}~\bibnamefont
  {Meloni}},\ }\href {\doibase 10.1007/JHEP08(2013)021} {\bibfield  {journal}
  {\bibinfo  {journal} {J.\ High Energy Phys.}\ }\textbf {\bibinfo {volume}
  {08}},\ \bibinfo {pages} {021} (\bibinfo {year} {2013})}\BibitemShut
  {NoStop}%
\bibitem [{\citenamefont {Dueck}\ and\ \citenamefont
  {Rodejohann}(2013)}]{Dueck:2013gca}%
  \BibitemOpen
  \bibfield  {author} {\bibinfo {author} {\bibfnamefont {A.}~\bibnamefont
  {Dueck}}\ and\ \bibinfo {author} {\bibfnamefont {W.}~\bibnamefont
  {Rodejohann}},\ }\href {\doibase 10.1007/JHEP09(2013)024} {\bibfield
  {journal} {\bibinfo  {journal} {J.\ High Energy Phys.}\ }\textbf {\bibinfo
  {volume} {09}},\ \bibinfo {pages} {024} (\bibinfo {year} {2013})}\BibitemShut
  {NoStop}%
\bibitem [{\citenamefont {Ohlsson}\ and\ \citenamefont
  {Pernow}(2019)}]{Ohlsson:2019sja}%
  \BibitemOpen
  \bibfield  {author} {\bibinfo {author} {\bibfnamefont {T.}~\bibnamefont
  {Ohlsson}}\ and\ \bibinfo {author} {\bibfnamefont {M.}~\bibnamefont
  {Pernow}},\ }\href {\doibase 10.1007/JHEP06(2019)085} {\bibfield  {journal}
  {\bibinfo  {journal} {J.\ High Energy Phys.}\ }\textbf {\bibinfo {volume}
  {06}},\ \bibinfo {pages} {085} (\bibinfo {year} {2019})}\BibitemShut
  {NoStop}%
\bibitem [{\citenamefont {Capozzi}\ \emph {et~al.}(2018)\citenamefont
  {Capozzi}, \citenamefont {Lisi}, \citenamefont {Marrone},\ and\ \citenamefont
  {Palazzo}}]{Capozzi:2018ubv}%
  \BibitemOpen
  \bibfield  {author} {\bibinfo {author} {\bibfnamefont {F.}~\bibnamefont
  {Capozzi}}, \bibinfo {author} {\bibfnamefont {E.}~\bibnamefont {Lisi}},
  \bibinfo {author} {\bibfnamefont {A.}~\bibnamefont {Marrone}}, \ and\
  \bibinfo {author} {\bibfnamefont {A.}~\bibnamefont {Palazzo}},\ }\href
  {\doibase 10.1016/j.ppnp.2018.05.005} {\bibfield  {journal} {\bibinfo
  {journal} {Prog.\ Part.\ Nucl.\ Phys.}\ }\textbf {\bibinfo {volume} {102}},\
  \bibinfo {pages} {48} (\bibinfo {year} {2018})}\BibitemShut {NoStop}%
\bibitem [{\citenamefont {de~Salas}\ \emph {et~al.}(2018)\citenamefont
  {de~Salas}, \citenamefont {Forero}, \citenamefont {Ternes}, \citenamefont
  {Tortola},\ and\ \citenamefont {Valle}}]{deSalas:2017kay}%
  \BibitemOpen
  \bibfield  {author} {\bibinfo {author} {\bibfnamefont {P.~F.}\ \bibnamefont
  {de~Salas}}, \bibinfo {author} {\bibfnamefont {D.~V.}\ \bibnamefont
  {Forero}}, \bibinfo {author} {\bibfnamefont {C.~A.}\ \bibnamefont {Ternes}},
  \bibinfo {author} {\bibfnamefont {M.}~\bibnamefont {Tortola}}, \ and\
  \bibinfo {author} {\bibfnamefont {J.~W.~F.}\ \bibnamefont {Valle}},\ }\href
  {\doibase 10.1016/j.physletb.2018.06.019} {\bibfield  {journal} {\bibinfo
  {journal} {Phys.\ Lett.\ B}\ }\textbf {\bibinfo {volume} {782}},\ \bibinfo
  {pages} {633} (\bibinfo {year} {2018})}\BibitemShut {NoStop}%
\bibitem [{\citenamefont {Esteban}\ \emph {et~al.}(2019)\citenamefont
  {Esteban}, \citenamefont {Gonzalez-Garcia}, \citenamefont
  {Hernandez-Cabezudo}, \citenamefont {Maltoni},\ and\ \citenamefont
  {Schwetz}}]{Esteban:2018azc}%
  \BibitemOpen
  \bibfield  {author} {\bibinfo {author} {\bibfnamefont {I.}~\bibnamefont
  {Esteban}}, \bibinfo {author} {\bibfnamefont {M.~C.}\ \bibnamefont
  {Gonzalez-Garcia}}, \bibinfo {author} {\bibfnamefont {A.}~\bibnamefont
  {Hernandez-Cabezudo}}, \bibinfo {author} {\bibfnamefont {M.}~\bibnamefont
  {Maltoni}}, \ and\ \bibinfo {author} {\bibfnamefont {T.}~\bibnamefont
  {Schwetz}},\ }\href {\doibase 10.1007/JHEP01(2019)106} {\bibfield  {journal}
  {\bibinfo  {journal} {J.\ High Energy Phys.}\ }\textbf {\bibinfo {volume}
  {01}},\ \bibinfo {pages} {106} (\bibinfo {year} {2019})}\BibitemShut
  {NoStop}%
\bibitem [{\citenamefont {Ade}\ \emph {et~al.}(2016)\citenamefont {Ade} \emph
  {et~al.}}]{Ade:2015xua}%
  \BibitemOpen
  \bibfield  {author} {\bibinfo {author} {\bibfnamefont {P.~A.~R.}\
  \bibnamefont {Ade}} \emph {et~al.} (\bibinfo {collaboration} {Planck
  Collaboration}),\ }\href {\doibase 10.1051/0004-6361/201525830} {\bibfield
  {journal} {\bibinfo  {journal} {Astron.\ Astrophys.}\ }\textbf {\bibinfo
  {volume} {594}},\ \bibinfo {pages} {A13} (\bibinfo {year}
  {2016})}\BibitemShut {NoStop}%
\bibitem [{\citenamefont {Palanque-Delabrouille}\ \emph
  {et~al.}(2015)\citenamefont {Palanque-Delabrouille} \emph
  {et~al.}}]{Palanque-Delabrouille:2015pga}%
  \BibitemOpen
  \bibfield  {author} {\bibinfo {author} {\bibfnamefont {N.}~\bibnamefont
  {Palanque-Delabrouille}} \emph {et~al.},\ }\href {\doibase
  10.1088/1475-7516/2015/11/011} {\bibfield  {journal} {\bibinfo  {journal}
  {J.\ Cosmol.\ Astropart.\ Phys.}\ }\textbf {\bibinfo {volume} {11}},\
  \bibinfo {pages} {011} (\bibinfo {year} {2015})}\BibitemShut {NoStop}%
\bibitem [{\citenamefont {Di~Valentino}\ \emph {et~al.}(2016)\citenamefont
  {Di~Valentino}, \citenamefont {Giusarma}, \citenamefont {Mena}, \citenamefont
  {Melchiorri},\ and\ \citenamefont {Silk}}]{DiValentino:2015sam}%
  \BibitemOpen
  \bibfield  {author} {\bibinfo {author} {\bibfnamefont {E.}~\bibnamefont
  {Di~Valentino}}, \bibinfo {author} {\bibfnamefont {E.}~\bibnamefont
  {Giusarma}}, \bibinfo {author} {\bibfnamefont {O.}~\bibnamefont {Mena}},
  \bibinfo {author} {\bibfnamefont {A.}~\bibnamefont {Melchiorri}}, \ and\
  \bibinfo {author} {\bibfnamefont {J.}~\bibnamefont {Silk}},\ }\href {\doibase
  10.1103/PhysRevD.93.083527} {\bibfield  {journal} {\bibinfo  {journal}
  {Phys.\ Rev.\ D}\ }\textbf {\bibinfo {volume} {93}},\ \bibinfo {pages}
  {083527} (\bibinfo {year} {2016})}\BibitemShut {NoStop}%
\bibitem [{\citenamefont {Zhang}(2016)}]{Zhang:2015uhk}%
  \BibitemOpen
  \bibfield  {author} {\bibinfo {author} {\bibfnamefont {X.}~\bibnamefont
  {Zhang}},\ }\href {\doibase 10.1103/PhysRevD.93.083011} {\bibfield  {journal}
  {\bibinfo  {journal} {Phys.\ Rev.\ D}\ }\textbf {\bibinfo {volume} {93}},\
  \bibinfo {pages} {083011} (\bibinfo {year} {2016})}\BibitemShut {NoStop}%
\bibitem [{\citenamefont {Cuesta}\ \emph {et~al.}(2016)\citenamefont {Cuesta},
  \citenamefont {Niro},\ and\ \citenamefont {Verde}}]{Cuesta:2015iho}%
  \BibitemOpen
  \bibfield  {author} {\bibinfo {author} {\bibfnamefont {A.~J.}\ \bibnamefont
  {Cuesta}}, \bibinfo {author} {\bibfnamefont {V.}~\bibnamefont {Niro}}, \ and\
  \bibinfo {author} {\bibfnamefont {L.}~\bibnamefont {Verde}},\ }\href
  {\doibase 10.1016/j.dark.2016.04.005} {\bibfield  {journal} {\bibinfo
  {journal} {Phys.\ Dark Univ.}\ }\textbf {\bibinfo {volume} {13}},\ \bibinfo
  {pages} {77} (\bibinfo {year} {2016})}\BibitemShut {NoStop}%
\bibitem [{\citenamefont {Giusarma}\ \emph {et~al.}(2016)\citenamefont
  {Giusarma}, \citenamefont {Gerbino}, \citenamefont {Mena}, \citenamefont
  {Vagnozzi}, \citenamefont {Ho},\ and\ \citenamefont
  {Freese}}]{Giusarma:2016phn}%
  \BibitemOpen
  \bibfield  {author} {\bibinfo {author} {\bibfnamefont {E.}~\bibnamefont
  {Giusarma}}, \bibinfo {author} {\bibfnamefont {M.}~\bibnamefont {Gerbino}},
  \bibinfo {author} {\bibfnamefont {O.}~\bibnamefont {Mena}}, \bibinfo {author}
  {\bibfnamefont {S.}~\bibnamefont {Vagnozzi}}, \bibinfo {author}
  {\bibfnamefont {S.}~\bibnamefont {Ho}}, \ and\ \bibinfo {author}
  {\bibfnamefont {K.}~\bibnamefont {Freese}},\ }\href {\doibase
  10.1103/PhysRevD.94.083522} {\bibfield  {journal} {\bibinfo  {journal}
  {Phys.\ Rev.\ D}\ }\textbf {\bibinfo {volume} {94}},\ \bibinfo {pages}
  {083522} (\bibinfo {year} {2016})}\BibitemShut {NoStop}%
\bibitem [{\citenamefont {Y\`eche}\ \emph {et~al.}(2017)\citenamefont
  {Y\`eche}, \citenamefont {Palanque-Delabrouille}, \citenamefont {Baur},\ and\
  \citenamefont {du~Mas~des Bourboux}}]{Yeche:2017upn}%
  \BibitemOpen
  \bibfield  {author} {\bibinfo {author} {\bibfnamefont {C.}~\bibnamefont
  {Y\`eche}}, \bibinfo {author} {\bibfnamefont {N.}~\bibnamefont
  {Palanque-Delabrouille}}, \bibinfo {author} {\bibfnamefont {J.}~\bibnamefont
  {Baur}}, \ and\ \bibinfo {author} {\bibfnamefont {H.}~\bibnamefont
  {du~Mas~des Bourboux}},\ }\href {\doibase 10.1088/1475-7516/2017/06/047}
  {\bibfield  {journal} {\bibinfo  {journal} {J.\ Cosmol.\ Astropart.\ Phys.}\
  }\textbf {\bibinfo {volume} {06}},\ \bibinfo {pages} {047} (\bibinfo {year}
  {2017})}\BibitemShut {NoStop}%
\bibitem [{\citenamefont {Aghanim}\ \emph {et~al.}(2018)\citenamefont {Aghanim}
  \emph {et~al.}}]{Aghanim:2018eyx}%
  \BibitemOpen
  \bibfield  {author} {\bibinfo {author} {\bibfnamefont {N.}~\bibnamefont
  {Aghanim}} \emph {et~al.} (\bibinfo {collaboration} {Planck Collaboration}),\
  }\href@noop {} {\  (\bibinfo {year} {2018})},\ \Eprint
  {http://arxiv.org/abs/1807.06209} {arXiv:1807.06209 [astro-ph.CO]}
  \BibitemShut {NoStop}%
%%CITATION = ARXIV:1807.06209;%%
\bibitem [{\citenamefont {Roy~Choudhury}\ and\ \citenamefont
  {Hannestad}(2019)}]{RoyChoudhury:2019hls}%
  \BibitemOpen
  \bibfield  {author} {\bibinfo {author} {\bibfnamefont {S.}~\bibnamefont
  {Roy~Choudhury}}\ and\ \bibinfo {author} {\bibfnamefont {S.}~\bibnamefont
  {Hannestad}},\ }\href@noop {} {\  (\bibinfo {year} {2019})},\ \Eprint
  {http://arxiv.org/abs/1907.12598} {arXiv:1907.12598 [astro-ph.CO]}
  \BibitemShut {NoStop}%
%%CITATION = ARXIV:1907.12598;%%
\bibitem [{\citenamefont {Dell'Oro}\ \emph {et~al.}(2015)\citenamefont
  {Dell'Oro}, \citenamefont {Marcocci}, \citenamefont {Viel},\ and\
  \citenamefont {Vissani}}]{Dell'Oro:2015tia}%
  \BibitemOpen
  \bibfield  {author} {\bibinfo {author} {\bibfnamefont {S.}~\bibnamefont
  {Dell'Oro}}, \bibinfo {author} {\bibfnamefont {S.}~\bibnamefont {Marcocci}},
  \bibinfo {author} {\bibfnamefont {M.}~\bibnamefont {Viel}}, \ and\ \bibinfo
  {author} {\bibfnamefont {F.}~\bibnamefont {Vissani}},\ }\href {\doibase
  10.1088/1475-7516/2015/12/023} {\bibfield  {journal} {\bibinfo  {journal}
  {J.\ Cosmol.\ Astropart.\ Phys.}\ }\textbf {\bibinfo {volume} {12}},\
  \bibinfo {pages} {023} (\bibinfo {year} {2015})}\BibitemShut {NoStop}%
%%CITATION = ARXIV:1505.02722;%%
\bibitem [{\citenamefont {Dell'Oro}\ \emph {et~al.}(2016)\citenamefont
  {Dell'Oro}, \citenamefont {Marcocci}, \citenamefont {Viel},\ and\
  \citenamefont {Vissani}}]{Dell'Oro:2016dbc}%
  \BibitemOpen
  \bibfield  {author} {\bibinfo {author} {\bibfnamefont {S.}~\bibnamefont
  {Dell'Oro}}, \bibinfo {author} {\bibfnamefont {S.}~\bibnamefont {Marcocci}},
  \bibinfo {author} {\bibfnamefont {M.}~\bibnamefont {Viel}}, \ and\ \bibinfo
  {author} {\bibfnamefont {F.}~\bibnamefont {Vissani}},\ }\href {\doibase
  10.1155/2016/2162659} {\bibfield  {journal} {\bibinfo  {journal} {Adv.\ High
  Energy Phys.}\ }\textbf {\bibinfo {volume} {2016}},\ \bibinfo {pages}
  {2162659} (\bibinfo {year} {2016})}\BibitemShut {NoStop}%
%%CITATION = ARXIV:1601.07512;%%
\bibitem [{\citenamefont {Vissani}(1999)}]{Vissani:1999tu}%
  \BibitemOpen
  \bibfield  {author} {\bibinfo {author} {\bibfnamefont {F.}~\bibnamefont
  {Vissani}},\ }\href {\doibase 10.1088/1126-6708/1999/06/022} {\bibfield
  {journal} {\bibinfo  {journal} {J.\ High Energy Phys.}\ }\textbf {\bibinfo
  {volume} {9906}},\ \bibinfo {pages} {022} (\bibinfo {year}
  {1999})}\BibitemShut {NoStop}%
%%CITATION = HEP-PH/9906525;%%
\bibitem [{\citenamefont {Benato}(2015)}]{Benato:2015via}%
  \BibitemOpen
  \bibfield  {author} {\bibinfo {author} {\bibfnamefont {G.}~\bibnamefont
  {Benato}},\ }\href {\doibase 10.1140/epjc/s10052-015-3802-1} {\bibfield
  {journal} {\bibinfo  {journal} {Eur.\ Phys.\ J.\ C}\ }\textbf {\bibinfo
  {volume} {75}},\ \bibinfo {pages} {563} (\bibinfo {year} {2015})}\BibitemShut
  {NoStop}%
\bibitem [{\citenamefont {Caldwell}\ \emph {et~al.}(2017)\citenamefont
  {Caldwell}, \citenamefont {Merle}, \citenamefont {Schulz},\ and\
  \citenamefont {Totzauer}}]{Caldwell:2017mqu}%
  \BibitemOpen
  \bibfield  {author} {\bibinfo {author} {\bibfnamefont {A.}~\bibnamefont
  {Caldwell}}, \bibinfo {author} {\bibfnamefont {A.}~\bibnamefont {Merle}},
  \bibinfo {author} {\bibfnamefont {O.}~\bibnamefont {Schulz}}, \ and\ \bibinfo
  {author} {\bibfnamefont {M.}~\bibnamefont {Totzauer}},\ }\href {\doibase
  10.1103/PhysRevD.96.073001} {\bibfield  {journal} {\bibinfo  {journal}
  {Phys.\ Rev.\ D}\ }\textbf {\bibinfo {volume} {96}},\ \bibinfo {pages}
  {073001} (\bibinfo {year} {2017})}\BibitemShut {NoStop}%
\bibitem [{\citenamefont {Agostini}\ \emph {et~al.}(2017)\citenamefont
  {Agostini}, \citenamefont {Benato},\ and\ \citenamefont
  {Detwiler}}]{Agostini:2017jim}%
  \BibitemOpen
  \bibfield  {author} {\bibinfo {author} {\bibfnamefont {M.}~\bibnamefont
  {Agostini}}, \bibinfo {author} {\bibfnamefont {G.}~\bibnamefont {Benato}}, \
  and\ \bibinfo {author} {\bibfnamefont {J.}~\bibnamefont {Detwiler}},\ }\href
  {\doibase 10.1103/PhysRevD.96.053001} {\bibfield  {journal} {\bibinfo
  {journal} {Phys.\ Rev.\ D}\ }\textbf {\bibinfo {volume} {96}},\ \bibinfo
  {pages} {053001} (\bibinfo {year} {2017})}\BibitemShut {NoStop}%
\bibitem [{\citenamefont {Pontecorvo}(1958)}]{Pontecorvo:1957qd}%
  \BibitemOpen
  \bibfield  {author} {\bibinfo {author} {\bibfnamefont {B.}~\bibnamefont
  {Pontecorvo}},\ }\href@noop {} {\bibfield  {journal} {\bibinfo  {journal}
  {Sov. Phys. JETP}\ }\textbf {\bibinfo {volume} {7}},\ \bibinfo {pages} {172}
  (\bibinfo {year} {1958})},\ \bibinfo {note} {[Zh.\ Eksp.\ Teor.\ Fiz.\ {\bf
  34}, 247 (1957)]}\BibitemShut {NoStop}%
%%CITATION = SPHJA,7,172;%%
\bibitem [{\citenamefont {Cirelli}\ \emph {et~al.}(2005)\citenamefont
  {Cirelli}, \citenamefont {Marandella}, \citenamefont {Strumia},\ and\
  \citenamefont {Vissani}}]{Cirelli:2004cz}%
  \BibitemOpen
  \bibfield  {author} {\bibinfo {author} {\bibfnamefont {M.}~\bibnamefont
  {Cirelli}}, \bibinfo {author} {\bibfnamefont {G.}~\bibnamefont {Marandella}},
  \bibinfo {author} {\bibfnamefont {A.}~\bibnamefont {Strumia}}, \ and\
  \bibinfo {author} {\bibfnamefont {F.}~\bibnamefont {Vissani}},\ }\href
  {\doibase 10.1016/j.nuclphysb.2004.11.056} {\bibfield  {journal} {\bibinfo
  {journal} {Nucl.\ Phys.\ B}\ }\textbf {\bibinfo {volume} {708}},\ \bibinfo
  {pages} {215} (\bibinfo {year} {2005})}\BibitemShut {NoStop}%
%%CITATION = HEP-PH/0403158;%%
\bibitem [{\citenamefont {Dentler}\ \emph {et~al.}(2018)\citenamefont
  {Dentler}, \citenamefont {Hern\'andez-Cabezudo}, \citenamefont {Kopp},
  \citenamefont {Machado}, \citenamefont {Maltoni}, \citenamefont
  {Martinez-Soler},\ and\ \citenamefont {Schwetz}}]{Dentler:2018sju}%
  \BibitemOpen
  \bibfield  {author} {\bibinfo {author} {\bibfnamefont {M.}~\bibnamefont
  {Dentler}}, \bibinfo {author} {\bibfnamefont {A.}~\bibnamefont
  {Hern\'andez-Cabezudo}}, \bibinfo {author} {\bibfnamefont {J.}~\bibnamefont
  {Kopp}}, \bibinfo {author} {\bibfnamefont {P.~A.~N.}\ \bibnamefont
  {Machado}}, \bibinfo {author} {\bibfnamefont {M.}~\bibnamefont {Maltoni}},
  \bibinfo {author} {\bibfnamefont {I.}~\bibnamefont {Martinez-Soler}}, \ and\
  \bibinfo {author} {\bibfnamefont {T.}~\bibnamefont {Schwetz}},\ }\href
  {\doibase 10.1007/JHEP08(2018)010} {\bibfield  {journal} {\bibinfo  {journal}
  {J.\ High Energy Phys.}\ }\textbf {\bibinfo {volume} {08}},\ \bibinfo {pages}
  {010} (\bibinfo {year} {2018})}\BibitemShut {NoStop}%
\bibitem [{\citenamefont {Diaz}\ \emph {et~al.}(2019)\citenamefont {Diaz},
  \citenamefont {Arg$\ddot{\text u}$elles}, \citenamefont {Collin},
  \citenamefont {Conrad},\ and\ \citenamefont {Shaevitz}}]{Diaz:2019fwt}%
  \BibitemOpen
  \bibfield  {author} {\bibinfo {author} {\bibfnamefont {A.}~\bibnamefont
  {Diaz}}, \bibinfo {author} {\bibfnamefont {C.~A.}\ \bibnamefont
  {Arg$\ddot{\text u}$elles}}, \bibinfo {author} {\bibfnamefont {G.~H.}\
  \bibnamefont {Collin}}, \bibinfo {author} {\bibfnamefont {J.~M.}\
  \bibnamefont {Conrad}}, \ and\ \bibinfo {author} {\bibfnamefont {M.~H.}\
  \bibnamefont {Shaevitz}},\ }\href@noop {} {\  (\bibinfo {year} {2019})},\
  \Eprint {http://arxiv.org/abs/1906.00045} {arXiv:1906.00045 [hep-ex]}
  \BibitemShut {NoStop}%
%%CITATION = ARXIV:1906.00045;%%
\bibitem [{\citenamefont {Girardi}\ \emph {et~al.}(2013)\citenamefont
  {Girardi}, \citenamefont {Meroni},\ and\ \citenamefont
  {Petcov}}]{Girardi:2013zra}%
  \BibitemOpen
  \bibfield  {author} {\bibinfo {author} {\bibfnamefont {I.}~\bibnamefont
  {Girardi}}, \bibinfo {author} {\bibfnamefont {A.}~\bibnamefont {Meroni}}, \
  and\ \bibinfo {author} {\bibfnamefont {S.~T.}\ \bibnamefont {Petcov}},\
  }\href {\doibase 10.1007/JHEP11(2013)146} {\bibfield  {journal} {\bibinfo
  {journal} {J.\ High Energy Phys.}\ }\textbf {\bibinfo {volume} {1311}},\
  \bibinfo {pages} {146} (\bibinfo {year} {2013})}\BibitemShut {NoStop}%
%%CITATION = ARXIV:1308.5802;%%
\bibitem [{\citenamefont {Giunti}\ and\ \citenamefont
  {Zavanin}(2015)}]{Giunti:2015kza}%
  \BibitemOpen
  \bibfield  {author} {\bibinfo {author} {\bibfnamefont {C.}~\bibnamefont
  {Giunti}}\ and\ \bibinfo {author} {\bibfnamefont {E.~M.}\ \bibnamefont
  {Zavanin}},\ }\href {\doibase 10.1007/JHEP07(2015)171} {\bibfield  {journal}
  {\bibinfo  {journal} {J.\ High Energy Phys.}\ }\textbf {\bibinfo {volume}
  {07}},\ \bibinfo {pages} {171} (\bibinfo {year} {2015})}\BibitemShut
  {NoStop}%
\bibitem [{\citenamefont {Huang}\ and\ \citenamefont
  {Zhou}(2019)}]{Huang:2019qvq}%
  \BibitemOpen
  \bibfield  {author} {\bibinfo {author} {\bibfnamefont {G.~Y.}\ \bibnamefont
  {Huang}}\ and\ \bibinfo {author} {\bibfnamefont {S.}~\bibnamefont {Zhou}},\
  }\href {\doibase 10.1016/j.nuclphysb.2019.114691} {\bibfield  {journal}
  {\bibinfo  {journal} {Nucl.\ Phys.\ B}\ }\textbf {\bibinfo {volume} {945}},\
  \bibinfo {pages} {114691} (\bibinfo {year} {2019})}\BibitemShut {NoStop}%
\bibitem [{\citenamefont {Riess}\ \emph {et~al.}(2019)\citenamefont {Riess},
  \citenamefont {Casertano}, \citenamefont {Yuan}, \citenamefont {Macri},\ and\
  \citenamefont {Scolnic}}]{Riess:2019cxk}%
  \BibitemOpen
  \bibfield  {author} {\bibinfo {author} {\bibfnamefont {A.~G.}\ \bibnamefont
  {Riess}}, \bibinfo {author} {\bibfnamefont {S.}~\bibnamefont {Casertano}},
  \bibinfo {author} {\bibfnamefont {W.}~\bibnamefont {Yuan}}, \bibinfo {author}
  {\bibfnamefont {L.~M.}\ \bibnamefont {Macri}}, \ and\ \bibinfo {author}
  {\bibfnamefont {D.}~\bibnamefont {Scolnic}},\ }\href {\doibase
  10.3847/1538-4357/ab1422} {\bibfield  {journal} {\bibinfo  {journal}
  {Astrophys.\ J.}\ }\textbf {\bibinfo {volume} {876}},\ \bibinfo {pages} {85}
  (\bibinfo {year} {2019})}\BibitemShut {NoStop}%
\bibitem [{\citenamefont {Verde}\ \emph {et~al.}(2019)\citenamefont {Verde},
  \citenamefont {Treu},\ and\ \citenamefont {Riess}}]{Verde:2019ivm}%
  \BibitemOpen
  \bibfield  {author} {\bibinfo {author} {\bibfnamefont {L.}~\bibnamefont
  {Verde}}, \bibinfo {author} {\bibfnamefont {T.}~\bibnamefont {Treu}}, \ and\
  \bibinfo {author} {\bibfnamefont {A.~G.}\ \bibnamefont {Riess}}\ }(\bibinfo
  {year} {2019})\ \Eprint {http://arxiv.org/abs/1907.10625} {arXiv:1907.10625
  [astro-ph.CO]} \BibitemShut {NoStop}%
%%CITATION = ARXIV:1907.10625;%%
\bibitem [{\citenamefont {de~Holanda}\ and\ \citenamefont
  {Smirnov}(2011)}]{deHolanda:2010am}%
  \BibitemOpen
  \bibfield  {author} {\bibinfo {author} {\bibfnamefont {P.~C.}\ \bibnamefont
  {de~Holanda}}\ and\ \bibinfo {author} {\bibfnamefont {A.~Y.}\ \bibnamefont
  {Smirnov}},\ }\href {\doibase 10.1103/PhysRevD.83.113011} {\bibfield
  {journal} {\bibinfo  {journal} {Phys.\ Rev.\ D}\ }\textbf {\bibinfo {volume}
  {83}},\ \bibinfo {pages} {113011} (\bibinfo {year} {2011})}\BibitemShut
  {NoStop}%
\bibitem [{\citenamefont {Alvis}\ \emph {et~al.}(2019)\citenamefont {Alvis}
  \emph {et~al.}}]{Alvis:2019sil}%
  \BibitemOpen
  \bibfield  {author} {\bibinfo {author} {\bibfnamefont {S.~I.}\ \bibnamefont
  {Alvis}} \emph {et~al.} (\bibinfo {collaboration} {MAJORANA Collaboration}),\
  }\href {\doibase 10.1103/PhysRevC.100.025501} {\bibfield  {journal} {\bibinfo
   {journal} {Phys.\ Rev.\ C}\ }\textbf {\bibinfo {volume} {100}},\ \bibinfo
  {pages} {025501} (\bibinfo {year} {2019})}\BibitemShut {NoStop}%
\bibitem [{\citenamefont {Agostini}\ \emph {et~al.}(2018)\citenamefont
  {Agostini} \emph {et~al.}}]{Agostini:2018tnm}%
  \BibitemOpen
  \bibfield  {author} {\bibinfo {author} {\bibfnamefont {M.}~\bibnamefont
  {Agostini}} \emph {et~al.} (\bibinfo {collaboration} {GERDA Collaboration}),\
  }\href {\doibase 10.1103/PhysRevLett.120.132503} {\bibfield  {journal}
  {\bibinfo  {journal} {Phys.\ Rev.\ Lett.}\ }\textbf {\bibinfo {volume}
  {120}},\ \bibinfo {pages} {132503} (\bibinfo {year} {2018})}\BibitemShut
  {NoStop}%
\bibitem [{\citenamefont {Alduino}\ \emph {et~al.}(2018)\citenamefont {Alduino}
  \emph {et~al.}}]{Alduino:2017ehq}%
  \BibitemOpen
  \bibfield  {author} {\bibinfo {author} {\bibfnamefont {C.}~\bibnamefont
  {Alduino}} \emph {et~al.} (\bibinfo {collaboration} {CUORE Collaboration}),\
  }\href {\doibase 10.1103/PhysRevLett.120.132501} {\bibfield  {journal}
  {\bibinfo  {journal} {Phys.\ Rev.\ Lett.}\ }\textbf {\bibinfo {volume}
  {120}},\ \bibinfo {pages} {132501} (\bibinfo {year} {2018})}\BibitemShut
  {NoStop}%
\bibitem [{\citenamefont {Anton}\ \emph {et~al.}(2019)\citenamefont {Anton}
  \emph {et~al.}}]{Anton:2019wmi}%
  \BibitemOpen
  \bibfield  {author} {\bibinfo {author} {\bibfnamefont {G.}~\bibnamefont
  {Anton}} \emph {et~al.} (\bibinfo {collaboration} {EXO-200 Collaboration}),\
  }\href@noop {} {\  (\bibinfo {year} {2019})},\ \Eprint
  {http://arxiv.org/abs/1906.02723} {arXiv:1906.02723 [hep-ex]} \BibitemShut
  {NoStop}%
%%CITATION = ARXIV:1906.02723;%%
\bibitem [{\citenamefont {Gando}\ \emph {et~al.}(2016)\citenamefont {Gando}
  \emph {et~al.}}]{KamLAND-Zen:2016pfg}%
  \BibitemOpen
  \bibfield  {author} {\bibinfo {author} {\bibfnamefont {A.}~\bibnamefont
  {Gando}} \emph {et~al.} (\bibinfo {collaboration} {KamLAND-Zen
  Collaboration}),\ }\href {\doibase 10.1103/PhysRevLett.117.109903} {\bibfield
   {journal} {\bibinfo  {journal} {Phys.\ Rev.\ Lett.}\ }\textbf {\bibinfo
  {volume} {117}},\ \bibinfo {pages} {082503} (\bibinfo {year}
  {2016})}\BibitemShut {NoStop}%
\bibitem [{\citenamefont {Dell'Oro}\ \emph {et~al.}(2014)\citenamefont
  {Dell'Oro}, \citenamefont {Marcocci},\ and\ \citenamefont
  {Vissani}}]{Dell'Oro:2014yca}%
  \BibitemOpen
  \bibfield  {author} {\bibinfo {author} {\bibfnamefont {S.}~\bibnamefont
  {Dell'Oro}}, \bibinfo {author} {\bibfnamefont {S.}~\bibnamefont {Marcocci}},
  \ and\ \bibinfo {author} {\bibfnamefont {F.}~\bibnamefont {Vissani}},\ }\href
  {\doibase 10.1103/PhysRevD.90.033005} {\bibfield  {journal} {\bibinfo
  {journal} {Phys.\ Rev.\ D}\ }\textbf {\bibinfo {volume} {90}},\ \bibinfo
  {pages} {033005} (\bibinfo {year} {2014})}\BibitemShut {NoStop}%
%%CITATION = ARXIV:1404.2616;%%
\bibitem [{\citenamefont {Biller}(2013)}]{Biller:2013wua}%
  \BibitemOpen
  \bibfield  {author} {\bibinfo {author} {\bibfnamefont {S.~D.}\ \bibnamefont
  {Biller}},\ }\href {\doibase 10.1103/PhysRevD.87.071301} {\bibfield
  {journal} {\bibinfo  {journal} {Phys.\ Rev. D}\ }\textbf {\bibinfo {volume}
  {87}},\ \bibinfo {pages} {071301} (\bibinfo {year} {2013})}\BibitemShut
  {NoStop}%
%%CITATION = ARXIV:1306.5654;%%
\end{thebibliography}%

\end{document}